\documentclass[aps, prb, twocolumn, showpacs, superscriptaddress]{revtex4-1}

\usepackage{graphicx}
\usepackage{dsfont}
\usepackage{amsmath}
\usepackage{amssymb}
\usepackage{physics} 
\usepackage{hyperref}
\hypersetup{colorlinks=true,
	        final=true, 
	        linkcolor=blue, 
	        citecolor=blue,
	        filecolor=blue,
	        urlcolor=blue
}

\newcommand{\eF}{\varepsilon_{\mathrm{F}}}
\newcommand{\Sbin}{\mathrm{Sb}_{\mathrm{in}}}
\newcommand{\Sbout}{\mathrm{Sb}_{\mathrm{out}}}

\newcommand{\myred}[1]{\textcolor{red}{#1}}

\begin{document}
\title{Tuning the van Hove singularities in AV$_{3}$Sb$_{5}$ (A = K, Rb, Cs) via pressure and doping}
\author{Harrison LaBollita}
\affiliation{Department of Physics, Arizona State University, Tempe, AZ 85287, USA}
\author{Antia S. Botana}
\affiliation{Department of Physics, Arizona State University, Tempe, AZ 85287, USA}
\date{\today}

%%%%%%%%%%%%%%%%% ABSTRACT %%%%%%%%%%%%%%%%%%%%%%%
\begin{abstract}
We investigate the electronic structure of the new family of kagome metals AV$_{3}$Sb$_{5}$ (A = K, Rb, Cs) using first-principles calculations. We   analyze systematically the evolution of the van Hove singularities (vHss) across the entire family upon applied pressure and hole doping, specifically focusing on the two vHss closer to the Fermi energy. With pressure, these two saddle points shift away from the Fermi level. At the same time, the Fermi surface undergoes a large reconstruction with respect to the Sb bands while the V bands remain largely unchanged, pointing to the relevant role of the Sb atoms in the electronic structure of these materials. Upon hole doping, we find the opposite trend, where the saddle points move closer to the Fermi level for increasing dopings. All in all, we show how pressure and doping are indeed two mechanisms that can be used to tune the location of the two vHss closer to the Fermi level and can be exploited to tune different Fermi surface instabilities and associated orders. 
\end{abstract}

\maketitle

%%%%%%%%%%%%%%%%% INTRODUCTION %%%%%%%%%%%%%%%%%%%%%%%
\section{\label{sec:intro}Introduction}
The kagome lattice, with its corner-sharing triangles, represents an ideal playground to give rise to exotic phenomena: from charge density wave formation (CDW) \cite{Isakov2006, Guo2009, Kiesel2012, Kiesel2013, Wang2013}, to Weyl and Dirac semimetals \cite{Liu2018, Liu2019, Morali2019}, and  unconventional superconductivity \cite{Ko2009, Kiesel2012, Kiesel2013, Wang2013}. These phenomena arise from the inherent features of the electronic structure of the kagome lattice: flat bands across the Brillouin zone, Dirac crossings appearing at the corner (K), and van Hove singularities (vHss)  at the edge (M). 

\begin{figure}
	\centering
	\includegraphics[width=\columnwidth]{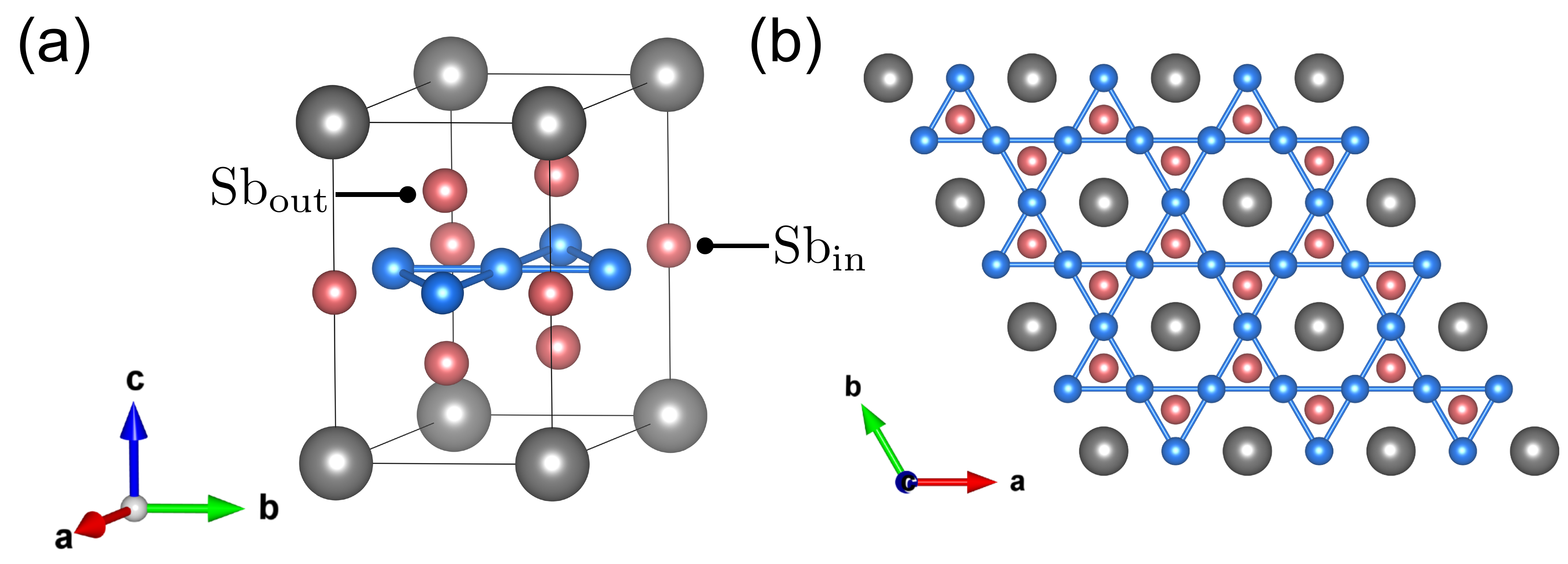}
	\caption{(a) Crystal structure of AV$_{3}$Sb$_{5}$ where the blue, pink, and grey spheres denote V, Sb, and A atoms, respectively. There are two inequivalent Sb atoms in the unit cell: $\Sbin$ which sits in the plane of the vanadium kagome net and $\Sbout$ which sits above and below the kagome plane. (b) Crystal structure in the $ab$ plane showing the vanadium kagome net.}
	\label{fig:struct}
\end{figure}

Recently, a new family of nonmagnetic kagome metals with chemical formula AV$_3$Sb$_5$ (A= K, Rb, Cs) have been discovered\cite{Ortiz2019new}.  Their crystal structure has P6/mmm symmetry displaying V$_3$Sb$_5$ slabs (with ideal kagome nets of V ions) and alternating A layers, as shown in Fig \ref{fig:struct}. All three systems exhibit a $\mathds{Z}_{2}$ topological band structure, superconductivity with a maximum transition temperature  T$_{c} \approx 0.9 - 2.5$ K, and CDW formation below T$_{\text{CDW}} \approx 78 - 103$ K \cite{ortiz2020z2Cs, ortiz2021z2K, ortiz2021super, Qiangwei2021}. This series of discoveries has triggered an immense amount of experimental and theoretical work \cite{Ortiz2019new, ortiz2020z2Cs, ortiz2021z2K, zhao2021cascade, zhao2021electronic, ortiz2021fermi, ortiz2021super, Kenney2021absence, Qiangwei2021, zhang2021pressure, kang2021twofold, hu2021rich, wu2021nature, lin2021complex, park2021electronic, cho2021emergence, Classen2020competing, hu2021chargeorderassisted, jiang2021discovery, song2021competition, song2021competing, wang2021competition, tsirlin2021anisotropic, shumiya2021tunable, uykur2021lowenergy, ratcliff2021coherent, li2021rotation, duan2021nodeless, uykur2021optical, du2021pressuretuned, wang2020proximityinduced, denner2021analysis, zhu2021doubledome, luo2021electronic, christensen2021theory, yin2021strainsensitive, ye2021flat, miao2021geometry, yang2019anomalous, luo2021distinct, qian2021revealing, Yu2021, setty2021electron, lou2021chargedensitywaveinduced, wang2021enhancement, wang2021unconventional, sun2021metalinsulator, nakayama2021multiple, Chen_2021, li2021indication, yu2021evidence}.  
AV$_3$Sb$_5$ materials display all the hallmarks of a kagome metal with nearly flat bands, linear crossings at K, and vHss at M \cite{Ortiz2019new, ortiz2020z2Cs, ortiz2021z2K, wu2021nature}. Earlier theoretical models on the kagome lattice predicted that both density wave order and superconductivity can arise at the vHs filling fractions \cite{Kiesel2012, Kiesel2013}. Based on this, the role of the multiple vHss in the vicinity of the Fermi level in AV$_3$Sb$_5$ has been deemed to be particularly crucial for the emergence (and competition) of the different Fermi surface instabilities \cite{kang2021twofold, hu2021rich, wu2021nature, lin2021complex, park2021electronic, cho2021emergence, Classen2020competing, hu2021chargeorderassisted, jiang2021discovery, song2021competition, song2021competing, wang2021competition}. As such, these materials represent an ideal platform to access and tune different orders via mechanisms like carrier doping or external pressure.

%%%%%%%%%%%%%%%  STRUCT FIG %%%%%%%%%%%%%%%

Hole doping, achieved through selective oxidation of exfoliated thin flakes, has indeed been shown to have a notable effect on both superconductivity and CDW order in the Cs variant \cite{song2021competition}. A superconducting dome is obtained as a function of doping content with a maximum T$_c$ = 4.7 K, significantly enhanced compared with the bulk. Pressure effects have recently been studied experimentally in all three compounds\cite{tsirlin2021anisotropic, zhu2021doubledome, du2021pressuretuned, Chen_2021}. Two superconducting domes arise upon applied pressure with enhanced T$_c$s and no sign of a structural phase transition \cite{du2021pressuretuned, zhu2021doubledome}. A recent study on the Cs variant \cite{tsirlin2021anisotropic} revealed a highly anisotropic compression due to the fast reduction of the Cs–Sb distances and suppression of Cs rattling motion. This prevents the Sb displacements required to stabilize the CDW state. These examples reflect the degree of tunability of these kagome materials upon carrier concentration and applied pressure, whose effects in the electronic structure remain to be investigated for the entire AV$_{3}$Sb$_{5}$ family. 

In this paper, using first principles calculations, we revisit the electronic structure of AV$_3$Sb$_5$ (A= K, Rb, Cs) and study its evolution upon applied pressure and hole doping in a systematic manner, with special focus on the two vHss closer to the Fermi level. Even though the known AV$_3$Sb$_5$ compounds are remarkably similar to each other, we find that the electronic structure of the Cs compound varies significantly with respect to the K and Rb materials. Upon applying external pressure, the Fermi surface of these materials undergoes a large reconstruction with respect to the Sb bands while the V bands remain essentially unchanged. Moreover, we find that the two saddle points move away from the Fermi energy in a linear fashion. Upon hole doping, we find the opposite trend, where the vHss move closer to the Fermi level when increasing the doping level. Overall, we  show how pressure and doping are indeed two mechanisms that can be used to tune the two vHss closer to the Fermi level and can be exploited to tune different Fermi surface instabilities and associated orders.

%%%%%%%%%%%%%%%%% METHODOLOGY %%%%%%%%%%%%%%%%%%%%%%%%
\section{\label{sec:method}Methodology}
 Density functional theory (DFT)-based calculations were performed using the all-electron, full potential code {\sc wien2k} based on the augmented plane wave plus local orbital (APW+lo) basis set \cite{w2k}. For the exchange-correlation functional, the Perdew-Burke-Ernzerhof (PBE) implementation of the generalized gradient approximation (GGA) was chosen \cite{pbe}.  Muffin-tin  radii of 2.50 a.u. for A (K, Rb, Cs) and V, and 2.43 a.u. for Sb were used, as well as a basis set cutoff  $\mathrm{RK}_{\text{max}} = 7$. All calculations presented in our manuscript have been run in the nonmagnetic state.  

\begin{table}
	\centering
	\begin{tabular*}{\columnwidth}{l@{\extracolsep{\fill}}rrrr}
	\hline 
	\hline
			                  & KV$_{3}$Sb$_{5}$ & RbV$_{3}$Sb$_{5}$ & CsV$_{3}$Sb$_{5}$  \\
		\hline
		$a$ (\AA{})            &   5.48          & 5.47              & 5.49	              \\
		$c$ (\AA{})            &   8.95          & 9.07	             & 9.31               \\
		$z$                    &   0.75          & 0.75              & 0.74               \\
		V-V (\AA{})            &   2.74		     & 2.74	             & 2.75	              \\
		V-$\Sbin$ (\AA{})      &   2.74		     & 2.74	             & 2.75	              \\
		V-$\Sbout$ (\AA{})     &   2.78		     & 2.77	             & 2.76	              \\
		$\theta$ ($^{\circ}$)  &  59.11	         & 59.11	         & 59.79	          \\
	\hline
	\hline 
	\end{tabular*}
	\caption{Experimental lattice parameters and relaxed atomic positions used in our calculations for AV$_{3}$Sb$_{5}$. Only the $z$ coordinate of the out-of-plane antimony atom ($\Sbout$) can change in the relaxations, while the other atoms sit at high symmetry positions ($z$ denotes the relaxed $z$ coordinate of the $\Sbout$ atoms). The derived $z$ coordinates agree with the experimentally reported ones: 0.75, 0.75, and 0.74 for K, Rb, and Cs, respectively \cite{Ortiz2019new}. Some relevant bond lengths (V-V, V-$\Sbin$, and V-$\Sbout$) are also shown. $\theta$ denotes the bond angle between V-$\Sbout$-V.}
	\label{tab:lattice}
\end{table}

Due to the variability in the position of the vHss with respect to the Fermi level across previous electronic structure calculations \cite{Ortiz2019new, ortiz2020z2Cs, ortiz2021z2K, ortiz2021fermi, zhao2021cascade, zhao2021electronic}, we have carefully checked the convergence of our calculations with respect to the size of the $k$-mesh. Very fine $k$-meshes ($\sim$ $10^{4}$ grid points in the full Brillouin zone) are needed to achieve convergence (see Appendix \ref{sec:kmesh} for more details). This highlights the importance of a proper convergence of DFT calculations with respect to the number of $k$-points in metallic systems like the ones we are dealing with here. Therefore, we used a very dense 38$\times$38$\times$20 ${\bf k}$-grid for integration in the irreducible Brillouin zone for our calculations. 

For the crystal structure (see Fig. \ref{fig:struct}), we relaxed the internal coordinates of all three materials within a nonmagnetic state, using the experimental lattice parameters and atomic positions as a starting point \cite{Ortiz2019new}. The lattice parameters and relaxed atomic positions used in our nonmagnetic electronic structure calculations are summarized in Table \ref{tab:lattice}. 

We have also used maximally-localized Wannier functions (MLWFs) to further investigate the electronic structure of the  AV$_{3}$Sb$_{5}$ family. To obtain the MLWFs, we employed {\sc wannier}90 \cite{wannier90} and {\sc wien}2{\sc wannier} \cite{wien2wannier}. Using the V-$d$, $\Sbin$-$p$, and $\Sbout$-$p$ orbitals for our initial projections, we obtained well-localized (albeit not unique) Wannier functions that correctly reproduce the band structure and orbital character (see Appendix \ref{sec:wannier} for more details).

For the calculations investigating pressure effects, we used the Vienna {\it ab-initio} Simulation Package (VASP) \cite{vasp} for the structural relaxations using projector augmented wave pseudopotentials \cite{PAW} with the GGA-PBE version of the exchange-correlation functional \cite{pbe}. For our plane-wave basis set, we used an energy cutoff of 300 eV for A = Cs, Rb and 350 eV for A = K. The same dense 38$\times$38$\times$20 ${\bf k}$-grid was used for the integration in the irreducible Brillouin zone. With the relaxed structures under applied external pressure, we performed our electronic structure calculations using the {\sc wien2k} code with the previously mentioned settings.

%%%%%%%%%%%%%% RESULTS %%%%%%%%%%%%%%%%%%%%%%%%%%%%%%%
\section{\label{sec:results}Results}

%%%%%%%%%%%%%%%%% ELECTRONIC STRUCTURE %%%%%%%%%%%%%%%
\subsection{\label{sec:electronic}Electronic structure of AV$_{3}$Sb$_{5}$ (A = K, Rb, Cs)}
%%%%%%%%%%%%%%% DFT FIG %%%%%%%%%%%%%%%
\begin{figure*}
	\centering
	\includegraphics[height=0.5\paperheight]{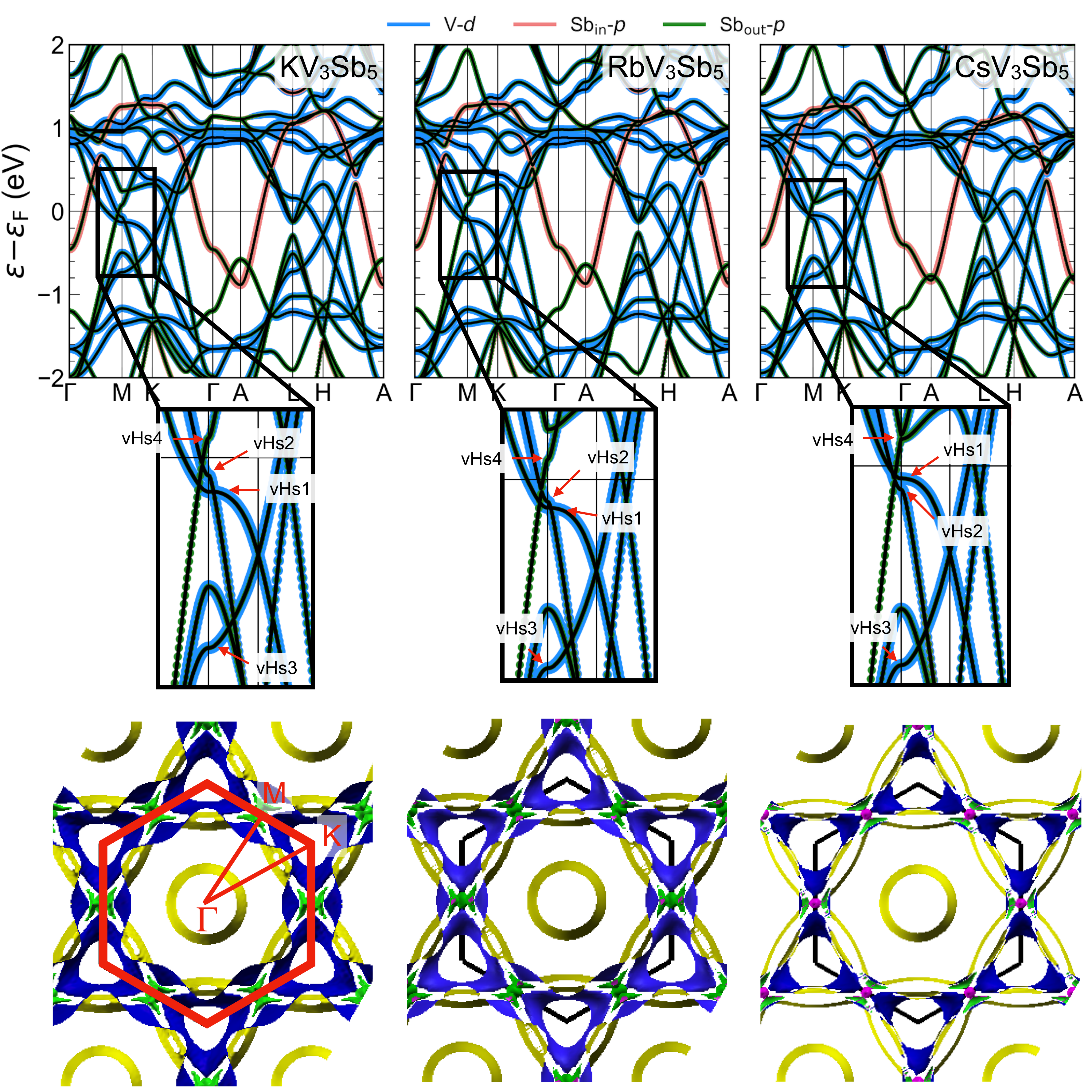}
	\caption{Electronic structure of AV$_{3}$Sb$_{5}$. Top: DFT band structures for A = K, Rb, Cs compounds (left to right) with fatbands indicating the V-$d$, Sb$_{\mathrm{in}}$-$p$, and Sb$_{\mathrm{out}}$-$p$ characters. Middle: Zoomed-in band structures around the M point highlighting the four vHss near $\eF$. Bottom: Corresponding Fermi surfaces for A = K, Rb, Cs compounds (left to right). In the left panel, the Brillouin zone is indicated in red and the high symmetry points for $k_z=0$ are labelled.}
	\label{fig:dft_electronic}
\end{figure*}

In Fig. \ref{fig:dft_electronic}, we summarize the nonmagnetic electronic structure of AV$_{3}$Sb$_{5}$ (A = K, Rb, Cs) showing the corresponding band structures and Fermi surfaces. Bands originating from the vanadium kagome net, as well as several bands of pure Sb-$5p$ origin can be observed in the vicinity of the Fermi level. Importantly, there are four vHss at the M point around the Fermi energy (labelled as vHs1-4 in Fig. \ref{fig:dft_electronic}), in agreement with previous work \cite{Ortiz2019new, ortiz2020z2Cs, ortiz2021z2K, tsirlin2021anisotropic}. The orbital content of vHs1 is mostly V-$d_{x^2-y^2}$, $d_{z^2}$, and $d_{xy}$ in character, whereas the other three vHss are dominated by V-$d_{xz}$ and $d_{yz}$ orbitals (see Appendix \ref{sec:es_cont_orb} Fig. \ref{fig:d_content}). The location of the two singularities closer to the Fermi level (vHs1 and vHs2) varies as the A cation is changed. We find that the flatter vHs1 (with $d_{z^{2}}$, $d_{x^2-y^2}$, $d_{xy}$ character) is located at  $-0.13$ eV, $-0.10$ eV, and $-0.05$ eV for K, Rb, and Cs, respectively. In turn, vHs2  (with $d_{xz}$, $d_{yz}$ character) is located at $-0.06$ eV, $-0.08$ eV, and $-0.1$ eV. Note that for the K and Rb compounds, vHs2 is located closer to $\eF$ than vHs1, whereas their positions are reversed in the Cs compound. 

Given that the different types of vHss around the Fermi level are likely responsible for the various Fermi surface instabilities in this family of materials\cite{kang2021twofold, hu2021rich}, an understanding of the interplay between the fermiology and the nearby vHss becomes important. As shown in Fig. \ref{fig:dft_electronic}, the fermiology of all three AV$_{3}$Sb$_{5}$ compounds  exhibits many similar features.  There are three distinct Fermi surface sheets in all cases: (i) a circular pocket around $\Gamma$ formed by $\Sbin$-$p_{z}$ orbitals, (ii) a hexagonal pocket with dominant V-$d_{xy}$, V-$d_{x^{2}-y^{2}}$, and V-$d_{z^{2}}$ character, and (iii) two triangular pockets composed of V-$d_{xz}/d_{yz}$ orbitals (see Ref. \onlinecite{luo2021electronic} for more details).  The (iii) triangular pocket (in proximity to vHs2) features a pronounced Fermi surface nesting so that it likely participates in the instability that causes CDW order \cite{hu2021rich, kang2021twofold, Classen2020competing}. In turn, the (ii) hexagonal pocket (in proximity to vHs1) has much weaker Fermi surface nesting. While the Fermi surfaces of the K, Rb, and Cs-based materials are qualitatively similar, there are subtle differences in the fermiology of the Cs compound compared to the K and Rb ones, which arise from the relative locations of vHs1 and vHs2. In the Cs compound, vHs1 lies closer to the Fermi energy than vHs2, as mentioned above, which alters the dispersion of these two bands near the Fermi energy, affecting the volume and shape of the $d_{xy}, d_{x^{2}-y^{2}}, d_{z^{2}}$ and $d_{xz}/d_{yz}$-related pockets. For the K and Cs compounds, we find excellent qualitative agreement with the Fermi surfaces from recent ARPES measurements \cite{luo2021electronic, kang2021twofold, hu2021rich}.

%%%%%%%%%%%%%%%%% WANNIER FUNCTION PERSPECTIVE %%%%%%%%%%%%%%%
We investigate further differences as the A cation changes (K, Rb, Cs) focusing on the degree of two-dimensionality (2D) of the electronic structure.  To this end, we calculate hopping integrals from maximally localized Wannier functions (see Appendix \ref{sec:wannier} for further details). Table \ref{tab:hoppings} shows all of the relevant hopping integrals obtained between the V-$d$, $\Sbin$-$p$, and $\Sbout$-$p$ orbitals. While there are sizeable out-of-plane hoppings from the kagome net to out-of-plane Sb atoms, we find that the majority of the hoppings are in-plane, pointing to the 2D-like character of the electronic structure of these materials. The most relevant hopping integrals in the K and Rb compounds are identical, both in terms of the involved orbitals and the size of the derived hoppings. For the Cs compound, there are fewer sizeable hoppings and the relevant hopping channels are different relative to the K and Rb compounds. Importantly, we find that the Cs compound exhibits fewer out-of-plane hoppings indicating that the Cs material is more 2D-like than the other two compounds. Given that the symmetry of the crystal structure of these materials is identical and that the position of the A states in the electronic structure does not get altered as A changes (see Appendix \ref{sec:es_cont_orb} Fig. \ref{fig:dos}), the slightly different bond angles and bond lengths between neighboring V, $\Sbin$, and $\Sbout$ atoms in the Cs compound (see Table \ref{tab:lattice})  could be at the origin of these differences. Our findings are in agreement with resistivity measurements that suggest that, even though all of these kagome materials are 2D metals, the Cs compound is more 2D-like \cite{ortiz2020z2Cs} as it displays the largest transport anisotropy.

%%%%%%%%%%%%%%%%% EFFECTS OF PRESSURE %%%%%%%%%%%%%%%
\subsection{\label{sec:pressure}Effects of pressure}
The pressure evolution (up to 20 GPa) of the crystal structure parameters for AV$_3$Sb$_5$ (A = K, Rb, Cs) is summarized in Fig. \ref{fig:vhs_press}(a) and in Appendix \ref{sec:es_press_dope} Fig. \ref{fig:press_distance}. The in-plane lattice parameter changes linearly and only slightly ($\sim$ 3\% reduction at 20 GPa), while the out-of-plane lattice parameter exhibits a much larger overall change ($\sim$ 15\% reduction at 20 GPa), with a low ($\lesssim 8.5$ GPa) and high  ($\gtrsim 8.5$ GPa) pressure regime where the out-of-plane lattice parameter changes at different rates. The $c/a$ ratio is systematically reduced under pressure in all systems by approximately the same amount (see Fig. \ref{fig:vhs_press}(a)). Applying pressure  only has a minor effect on the nearest-neighbor V-Sb distance (for both in- and out-of-plane Sb atoms), while there is a much larger change in the A-$\Sbout$ distance (see Appendix \ref{sec:es_press_dope} Fig. \ref{fig:press_distance}). The evolution of the A-$\Sbout$ distance with pressure compared to the V-V, V-$\Sbin$, V-$\Sbout$ bond length evolution agrees with experimental data for the Cs compound \cite{tsirlin2021anisotropic}. This scenario supports the picture that the V$_3$Sb$_5$ slabs are a rigid structural unit that is weakly coupled to the interstitial A layers \cite{tsirlin2021anisotropic}.

We focus now on tracking the change in energy (relative to the Fermi level) of the saddle points closer to it (vHs1 and vHs2). The shift that these vHss experience with respect to the Fermi level under external pressure is shown in Fig. \ref{fig:vhs_press}(b). In all three cases, both of these saddle points move away from the Fermi level (in a linear fashion) upon applied pressure. In addition to the two vHss, it is important to track the changes in the band structures. The evolution of the band structures with applied pressure (see Appendix \ref{sec:es_press_dope} Fig. \ref{fig:pressure}) reveals that the V bands are only slightly broadened, whereas the Sb bands undergo major changes. Specifically, the $\Sbout$- and $\Sbin$-$p_{z}$ bands between $\Gamma$ and A change drastically: the $\Sbout$  band that peaks at $\sim$ $-1$ eV at A at ambient pressure, gradually shifts up in energy, rising above $\eF$ between 10 and 15 GPa. This reconstruction of the bands around the Fermi level can be understood from the displacement of the $\Sbout$ atoms described above. These results show that even though the kagome bands (with their associated vHss) have been considered the only crucial ingredient for the electronic structure of AV$_{3}$Sb$_{5}$, the role of the Sb-$5p$ states cannot be disregarded, particularly upon applying pressure.

%%%%%%%%%%%%%%% PRESSURE FIG %%%%%%%%%%%%%%%
\begin{figure}
\centering
	\includegraphics[width=\columnwidth]{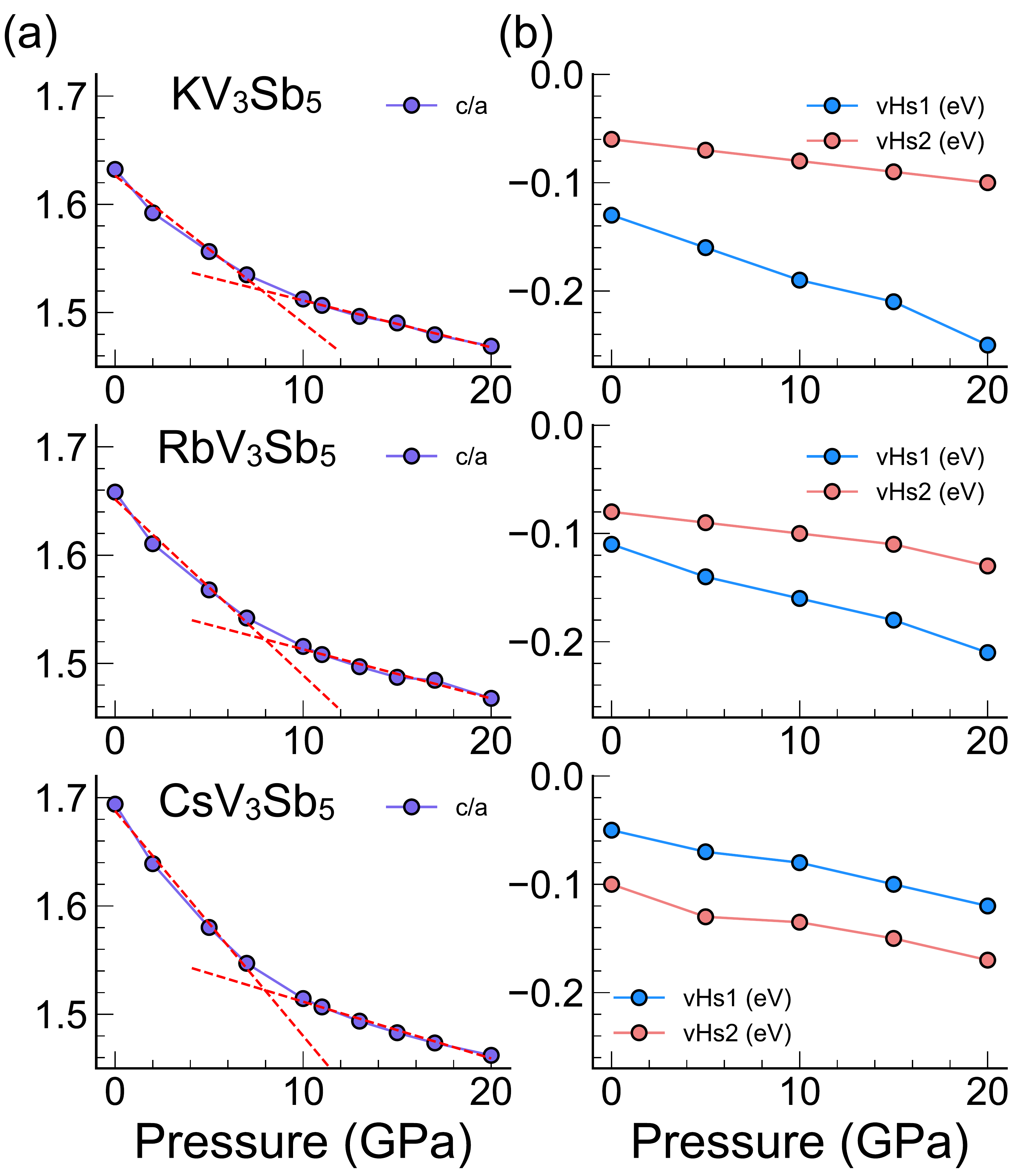}
	\caption{Evolution upon applied pressure of (a) the $c/a$ ratio with linear fits to the low and high pressure regimes and (b) the position of vHs1 and vHs2  relative to $\eF$ in AV$_{3}$Sb$_{5}$ (A= K, Rb, Cs).}
	\label{fig:vhs_press}
\end{figure}

Our findings are in agreement with the recently reported anisotropic compression in the Cs compound (with similar $c/a$ trends to those we report) due to the fast shrinkage of the Cs–Sb distances and suppression of Cs rattling motion\cite{tsirlin2021anisotropic}. We show here that this response can be extended to the K and Rb materials as well. Our results are also consistent with experimental data both in the context of superconductivity and CDW order. In the experimental temperature vs. pressure phase diagram, there are two superconducting domes for all three compounds \cite{zhang2021pressure, zhu2021doubledome}. For the Cs compound, the termination of the first superconducting dome corresponds to a kink in the $c/a$ ratio at $\sim$ 8 GPa \cite{zhang2021pressure}. From our calculated lattice parameters, we correctly capture a kink at the intersection between the low and high pressure regimes at $\sim$ 7.8, 7.9, and 8.0 GPa for A = K, Rb, and Cs, respectively (see Fig. \ref{fig:vhs_press}(a)). Furthermore, we capture the correct trend of the first superconducting dome terminating at higher pressure when increasing the A cation size \cite{zhang2021pressure, du2021pressuretuned, zhu2021doubledome}. The CDW state is suppressed at $\sim 1 - 2$ GPa in all materials \cite{du2021pressuretuned, wang2021competition, Chen_2021}. This suppression can be understood from the rapid decrease in the A-$\Sbout$ distance that makes the electronic structure more dispersive along the $c$-axis, weakening the nesting vector for the CDW order \cite{tsirlin2021anisotropic, wang2021competition, Chen_2021}. Given that the K and Rb compounds display a larger $c$-axis dispersion already at ambient pressure (see their larger out-of-plane hoppings, highlighted in the previous section), this scenario also supports the earlier suppression of CDW order upon applied pressure in these compounds when compared to their Cs counterpart.

%%%%%%%%%%%%%%%%% EFFECTS OF DOPING %%%%%%%%%%%%%%%
\subsection{\label{sec:dope}Effects of hole doping}
Another knob that can be used to tune the position of the vHss in the AV$_{3}$Sb$_{5}$ family is doping.  Experimentally, charge modulation in bulk samples of AV$_3$Sb$_5$ through chemical doping has not been realized. However, it has recently been shown that doping these materials in thin film form is possible via selective oxidation of exfoliated thin flakes \cite{song2021competition}: controlling the thickness of the flakes, the carrier concentration can be modulated and hole doping has been effectively achieved. To study the effects of hole doping theoretically, we employ the virtual crystal approximation (VCA), where we replace the A cation with an effective $\mathcal{A} = \mathrm{A} - x$ cation. We study a large range of dopings from $x = 0.0 - 0.3$ to capture general trends in the electronic structure.

\begin{figure}
    \centering
    \includegraphics[width=\columnwidth]{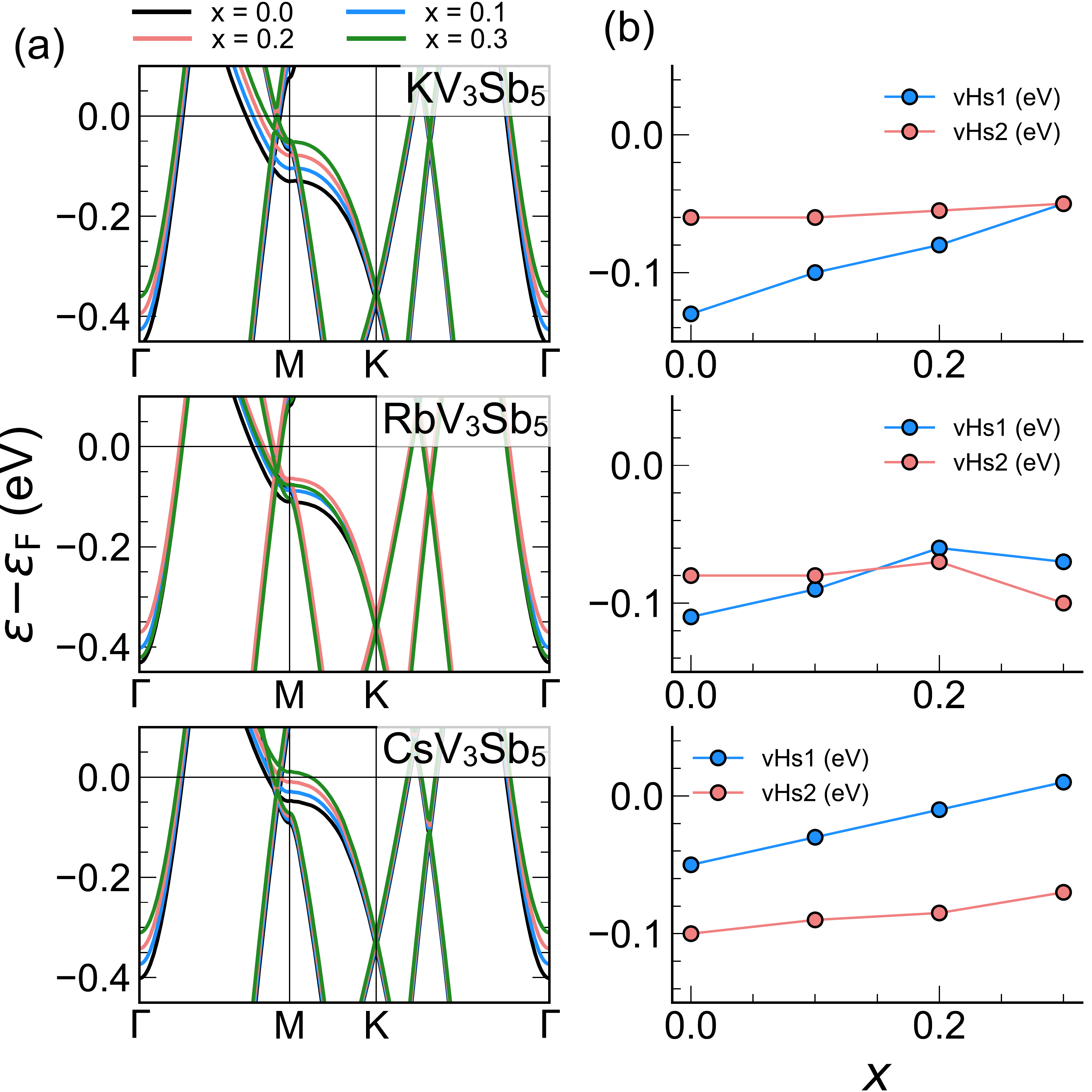}
    \caption{Evolution upon hole doping ($x$) within VCA of the (a) band structures and (b) position of vHs1 and vHs2 with respect to $\eF$ in AV$_{3}$Sb$_{5}$ (A= K, Rb, Cs).}
    \label{fig:dope_summ}
\end{figure}

From the band structures in Fig. \ref{fig:dope_summ}(a), we find that doping in AV$_3$Sb$_5$ (A = K, Rb, Cs) is highly orbitally-selective rather than giving rise to a simple rigid-band shift. The holes mainly dope the bands at $\Gamma$ (with dominant $\Sbin$-$p_{z}$ character) and the bands at M involved in vHs1 of $d_{x^2-y^2}$, $d_{z^2}$, and $d_{xy}$ character. Notably, the bands associated with vHs1 experience much larger shifts with doping compared to those associated with vHs2. This orbital-selective doping  further highlights the important role that the Sb states play in this family of kagome metals. 

%%%%%%%%%%%%%%% DOPE FIG %%%%%%%%%%%%%%%

In Fig. \ref{fig:dope_summ}(b), we track the relative position of the two vHss closer to $\eF$ (vHs1 and vHs2) upon hole doping. Across all materials, we find the opposite overall trend to the pressure case as the saddle points shift closer to the Fermi level with increasing doping levels (with applied pressure the saddle points moved away from it instead). Given that the doping is orbitally-selective, vHs1 is more sensitive to dopants than vHs2, as mentioned above. In the K compound, the vHss converge to nearly the same location at the highest doping, with vHs1 shifting up in energy significantly when compared to vHs2. Throughout all the previous sections, the K and Rb compounds have been shown to display an electronic structure that is practically identical, but upon doping we find a slightly different trend in the shifts of the two vHss. For the K compound, vHs2 remains essentially stationary, while vHs1 rapidly climbs in energy with doping. In the Rb compound,  vHs2 also remains essentially stationary at low dopings, however, at high dopings the vHss cross. This makes vHs1 now closer to the Fermi energy than vHs2, which would bring the fermiology of the Rb compound upon doping closer to the fermiology to the Cs compound. Finally, for the Cs compound, we find that vHs1 actually crosses the Fermi level and sits above it at the highest doping level considered, while vHs2 gradually shifts up in energy but with a much shallower slope. Additionally, for the Cs compound, we observe that the total density of states (DOS) at the Fermi level gradually increases with doping up to $x=0.2$  (see Appendix \ref{sec:es_press_dope} Fig. \ref{fig:cs_dos}). This trend in the DOS matches the increase in the superconducting critical temperature observed with doping in Ref. \onlinecite{song2021competition}. Overall, hole doping seems to be an effective way to shift the vHss in AV$_{3}$Sb$_{5}$ towards the Fermi level - we anticipate that it should have a larger effect in tuning vHs1.

%%%%%%%%%%%%%%%%% SUMMARY %%%%%%%%%%%%%%%
\section{\label{sec:conclusion}Summary}
We have used first-principles calculations to investigate the nonmagnetic electronic structure of the AV$_3$Sb$_5$ (A= K, Rb, Cs) family and tracked the evolution of the two vHss closer to the Fermi energy upon applied pressure and hole doping.  When applying external pressure, the two vHss  move away from the Fermi level and the Fermi surface undergoes a large reconstruction with respect to the Sb bands. This finding points to the role the Sb-$5p$ states play in the interesting physics exhibited by these materials. Upon hole doping, we find the opposite trend as the two vHss move closer to the Fermi level with increasing doping. Furthermore, doping is highly  orbitally-selective. Overall, we find that pressure and doping seem to be two effective mechanisms to tune the vHss closer to the Fermi level, whose role  has  been  deemed  to  be   crucial  for the emergence of the different Fermi surface instabilities in the AV$_3$Sb$_5$ family.

\section*{Acknowledgements}
 We acknowledge the support from NSF-DMR 2045826 and from the ASU Research Computing Center for HPC resources.

\bibliography{ref.bib}

%%%%%%%%%%%%%%%%% APPENDIX %%%%%%%%%%%%%%%

\clearpage
\newpage
\onecolumngrid % single column appendix

\appendix

\section{\label{sec:es_cont}Further details on the electronic structure calculations of AV$_3$Sb$_5$}

\subsection{\label{sec:kmesh} Convergence criteria of the DFT calculations}

\begin{figure}[ht]
    \centering
    \includegraphics[width=0.5\columnwidth]{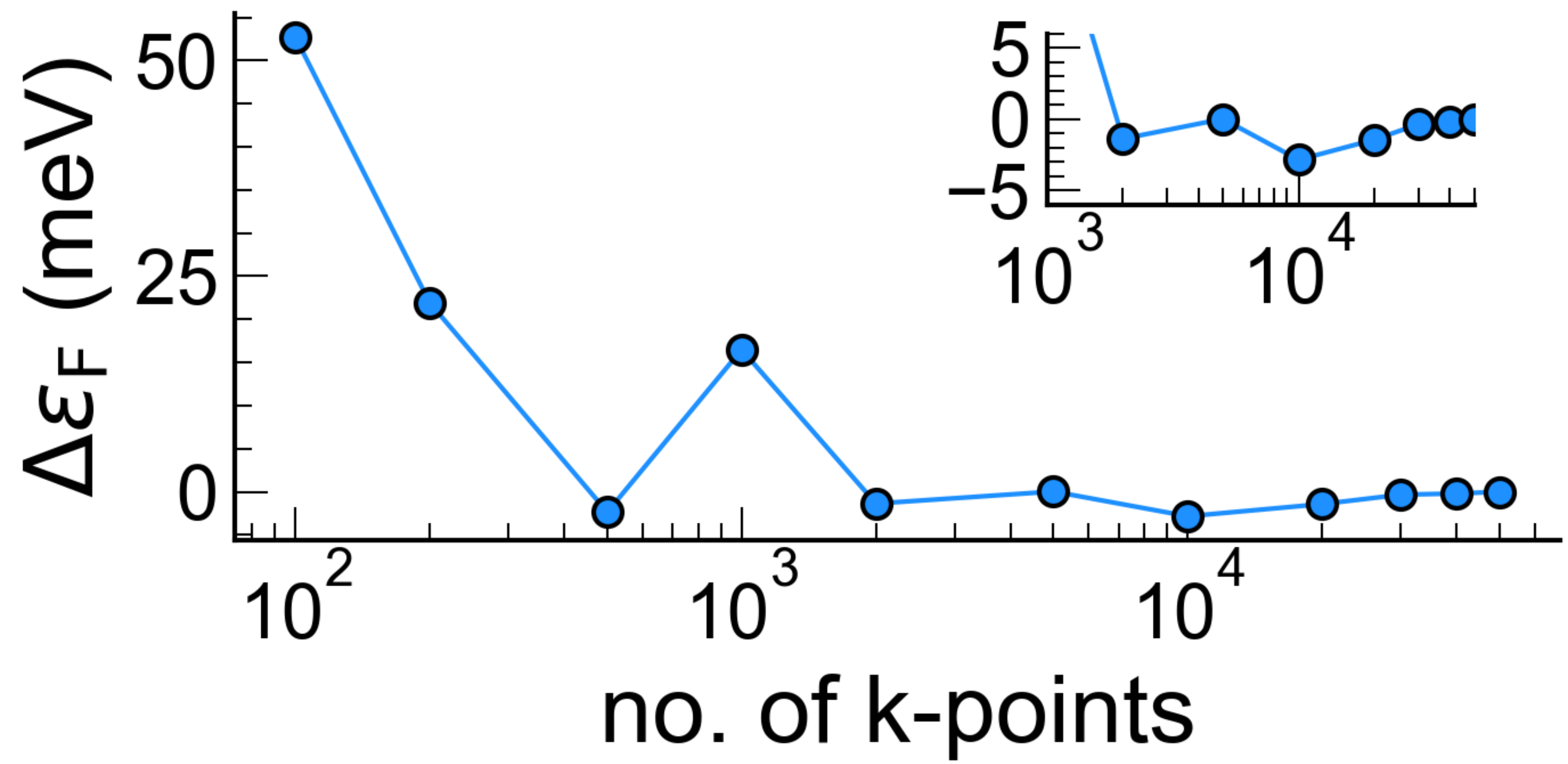}
    \caption{Convergence of the Fermi energy ($\varepsilon_{\text{F}}$) with respect to the number of $k$-points used in our grid. Here, $\Delta\varepsilon_{\mathrm{F}} = \varepsilon_{\mathrm{F}} - \varepsilon_{\mathrm{F}}^{\text{converged}}$. The inset highlights the convergence of the Fermi energy using a very dense mesh with more than 10$^{4}$ grid points}
    \label{fig:kmesh}
\end{figure}

Due to the variability in the position of the vHss with respect to the Fermi level across previous electronic structure calculations \cite{Ortiz2019new, ortiz2020z2Cs, ortiz2021z2K, ortiz2021fermi, zhao2021cascade, zhao2021electronic}, we carefully check the convergence of the Fermi energy ($\varepsilon_{\mathrm{F}}$) with respect to the size of the $k$-mesh. Because these systems are metallic with many bands crossing the Fermi level, a proper $k$-mesh is essential for accurately describing the electronic structure and, more importantly, the location of the vHss which are crucial to understanding the physics of these materials. In Fig. \ref{fig:kmesh}, we plot $\Delta \varepsilon_{\mathrm{F}}$ (where the converged Fermi energy is taken as a reference) versus the number of $k$-points used in our ${\bf k}$-grid. The inset in Fig. \ref{fig:kmesh} shows that the convergence of the Fermi energy to four decimal places occurs when the number of $k$-points is on the order of $10^{4}$ points in the full Brillouin zone. We have performed all of our electronic structure calculations throughout using this very dense mesh. 

\subsection{\label{sec:es_cont_orb}Orbital content and density of states}
\begin{figure*}[ht]
	\centering
	\includegraphics[height=0.33\paperheight]{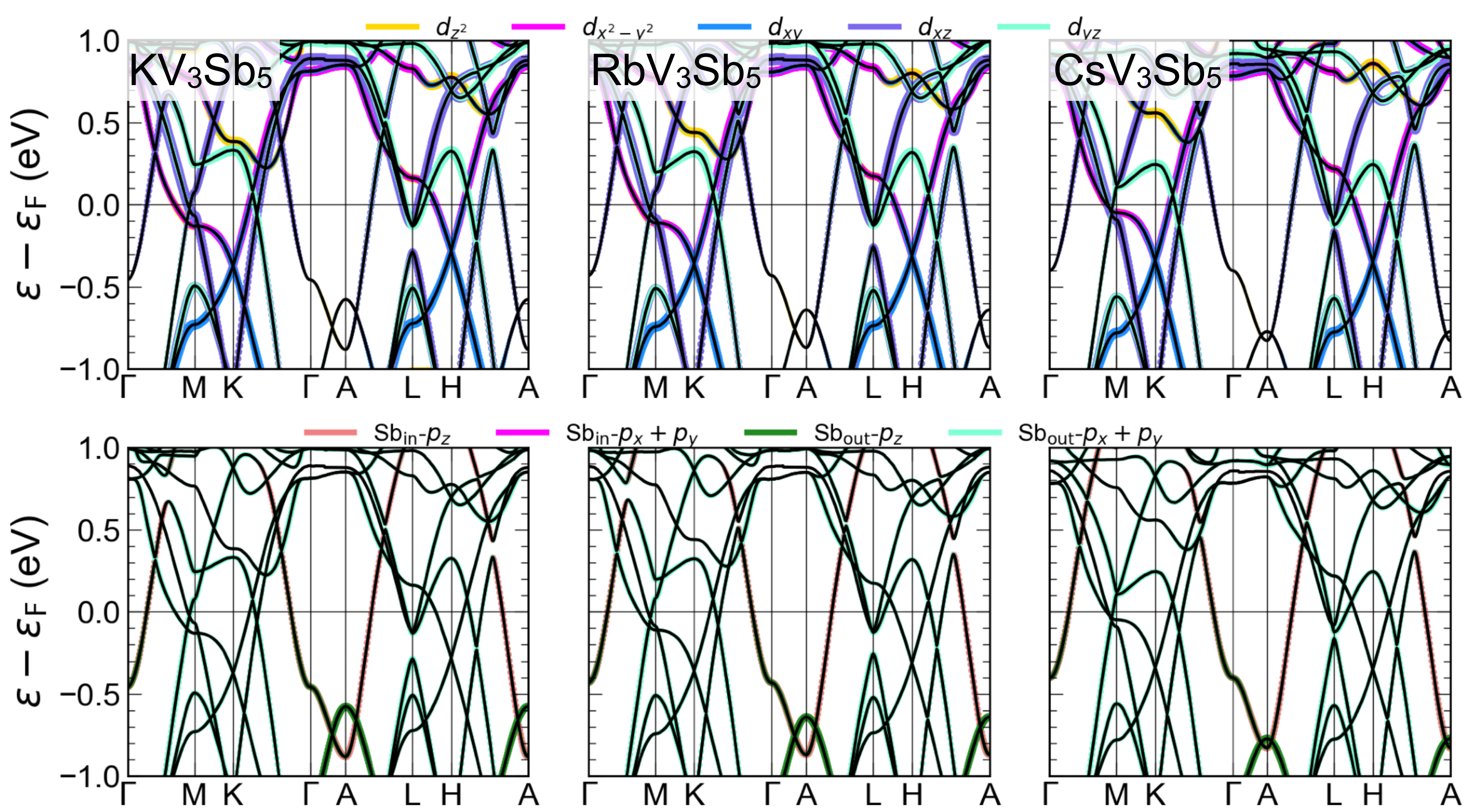}
	\caption{Nonmagnetic band structures with fatband representation for all of the relevant orbitals for AV$_{3}$Sb$_{5}$ (A = K, Rb, Cs) with A = K (left), A = Rb (middle) and A = Cs (right). Top: V-$3d$ orbitals. Bottom: $\Sbin$- and $\Sbout$-$p$ orbitals.}
	\label{fig:d_content}
\end{figure*}

\begin{figure}[ht]
    \centering
    \includegraphics[width=0.5\columnwidth]{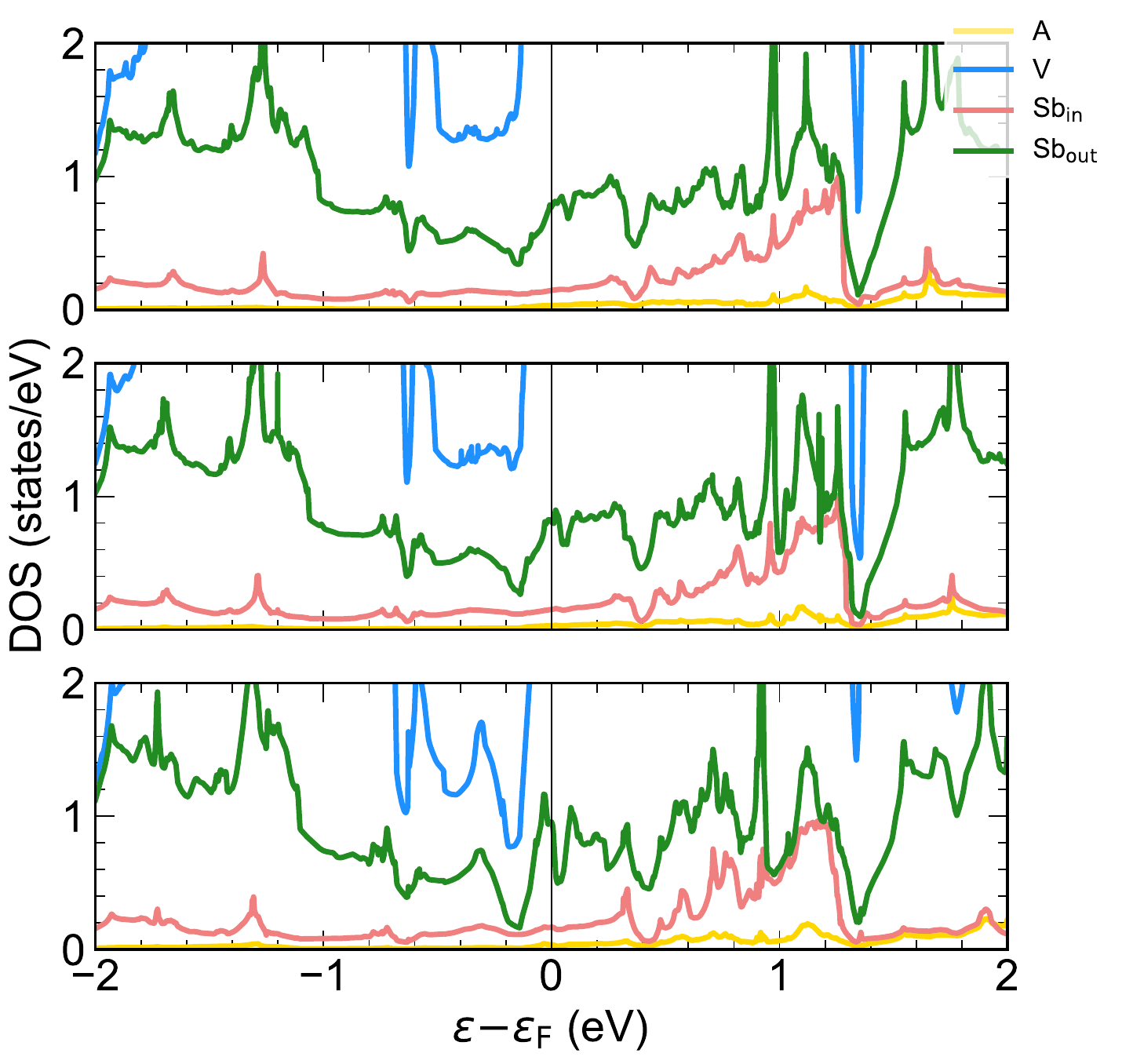}
    \caption{Atom-resolved density of states (DOS) for AV$_{3}$Sb$_{5}$ (A = K, Rb, Cs).}
    \label{fig:dos}
\end{figure}

In Fig. \ref{fig:d_content}, we further decompose the fatband representation of the band structures shown in Fig. \ref{fig:dft_electronic} into individual V-$d$ and Sb-$p$ orbitals. The top row shows the entire $d$-manifold, where the weight of the color around the band indicates the dominant orbital character of that band. We find that the bands comprising vHs1 are mainly $d_{xy}$, $d_{x^{2}-y^{2}}$, and $d_{z^{2}}$ character, while the bands comprising vHs2 are of $d_{xz}$ and $d_{yz}$ character. The bottom row breaks down the $p$ orbitals for both $\Sbin$ and $\Sbout$. 
To show that the A states do not play a role in the low energy physics of these systems, we plot the atom-resolved density of states (DOS) for all three compounds around the Fermi level in Fig. \ref{fig:dos}. We find that the A states (depicted in yellow) are essentially stagnant around the Fermi level, while the V states are dominant exhibiting a very large DOS characteristic of the vHss near the Fermi level.

\section{\label{sec:wannier}Wannier functions and hopping integrals}
\begin{figure}[ht]
    \centering
    \includegraphics[width=0.5\columnwidth]{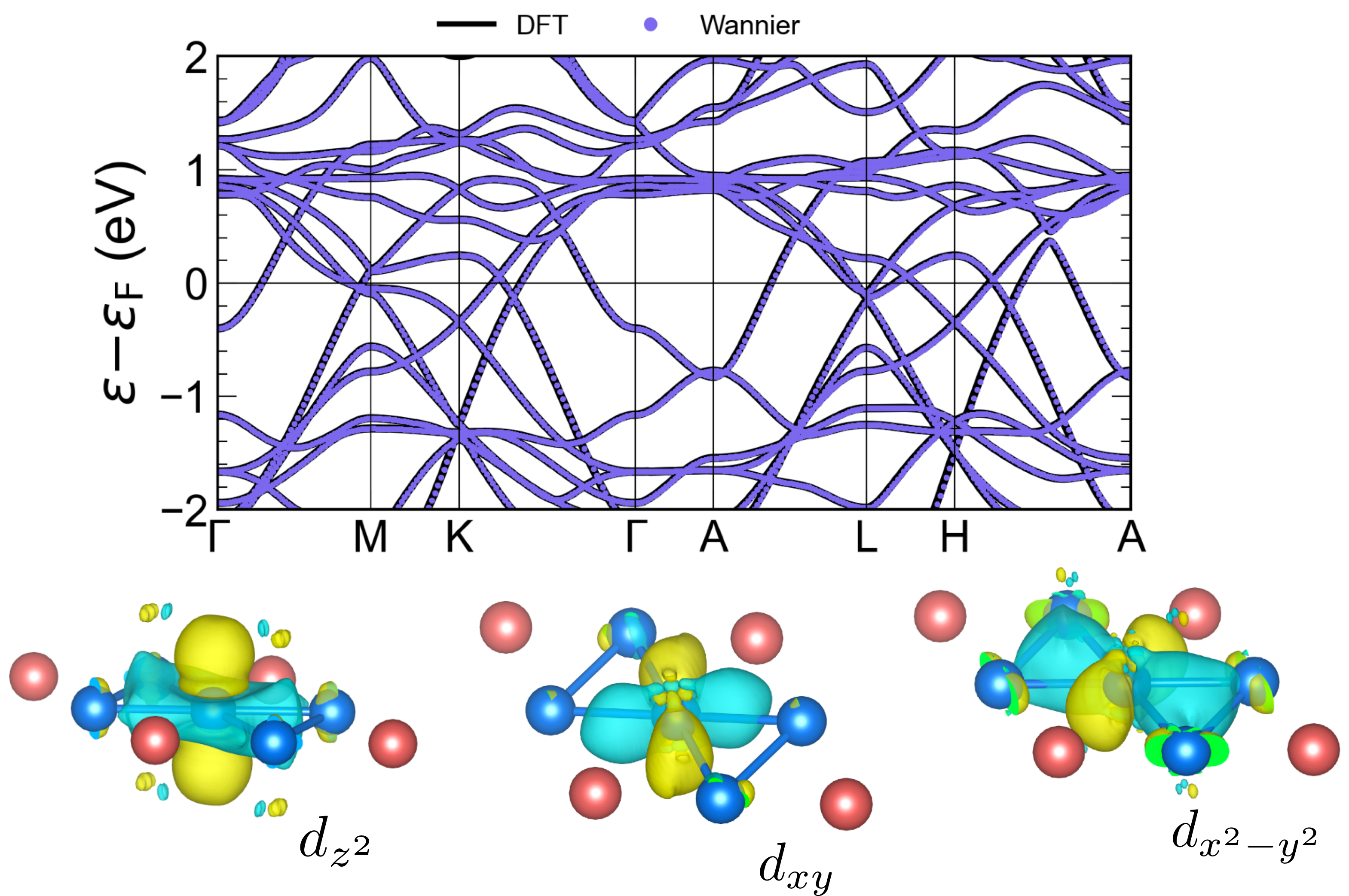}
    \caption{Summary of the wannierization for AV$_{3}$Sb$_{5}$. Here, we have taken A = Cs as an example. The fits of the other two materials are essentially identical. Top: DFT bands (black) compared to the Wannier function interpolation (light purple) along the high-symmetry path in the Brillouin zone. Bottom: Real-space representation of selected Wannier functions corresponding to the $d$ orbitals on the vanadium sites.}
    \label{fig:wann_bands}
\end{figure}

We analyze the local properties of the electronic structure of AV$_3$Sb$_5$ using MLWFs for all three compounds. In Fig. \ref{fig:wann_bands}, we have summarized the Wannier fits and some relevant Wannier functions. The agreement between the band structures obtained from the Wannier function interpolation and those derived from the DFT calculations is excellent, indicating a faithful (though not unique) transformation to Wannier functions. The fatband representation of our Wannier dispersion shows the orbital character of these bands describes $d$-like orbitals coming from the vanadium atoms and $p$-like orbitals coming from the  antimony atoms, in analogy to the DFT bands.
To investigate the degree of two-dimensionality of the electronic structure of these systems, we calculate hopping integrals from our Wannier functions and analyze the dominant hopping channels. We have summarized the largest hopping integrals in Table \ref{tab:hoppings}. We have used the notation
\begin{equation}
\langle \phi_{\alpha} | \mathcal{H} | \phi_{\beta} \rangle = t_{\alpha}^{\beta},
\end{equation}
where $\phi_{\alpha}$ and $\phi_{\beta}$ correspond to the Wannier function $\alpha, \beta \in $ [$d$, $p^{\mathrm{in}}$, $p^{\mathrm{out}}$] with $\alpha \neq \beta$ in our table and $\mathcal{H}$ denotes the Wannier Hamiltonian. $t_{\alpha}^{\beta}$ denotes the hopping integral describing the exchange from $\alpha$ to $\beta$. 

\begin{table*}
	\centering
	\begin{tabular*}{0.75\columnwidth}{l@{\extracolsep{\fill}}rlrlr}
	\hline
	\hline
	\multicolumn{2}{c}{KV$_{3}$Sb$_{5}$}    &  \multicolumn{2}{c}{RbV$_{3}$Sb${5}$}   & \multicolumn{2}{c}{CsV$_{3}$Sb$_{5}$} \\
	\hline
$ \langle d_{xy} |\mathcal{H}| d_{xy} \rangle $ 	              &  $0.62$  & $ \langle d_{xy} |\mathcal{H}| d_{xy}  \rangle $ & $0.63$							             &  $ \langle d_{xy} |\mathcal{H}| p_{y}^{\mathrm{in}} \rangle$ &  $-0.69$  \\
$ \langle d_{xy} |\mathcal{H}| p_{y}^{\mathrm{in}} \rangle $          &  $0.69$  & $ \langle d_{xy} |\mathcal{H}| p_{y}^{\mathrm{in}} \rangle $ & $0.70$				                     &  $ \langle d_{xz} |\mathcal{H}| p_{z}^{\mathrm{in}} \rangle$ &  $0.62$  \\
$ \langle d_{xz} |\mathcal{H}| p_{z}^{\mathrm{in}} \rangle $ 	      &  $0.62$  & $ \langle d_{xz} |\mathcal{H}| p_{z}^{\mathrm{in}} \rangle $ & $0.63$			                             &  $ \langle d_{x^{2}-y^{2}} |\mathcal{H}| p_{x}^{\mathrm{in}} \rangle$ &  $\myred{ 0.80 }$  \\
$ \langle d_{x^{2}-y^{2}} |\mathcal{H}| p_{x}^{\mathrm{in}} \rangle $ &  $\myred{ 0.80 }$  & $ \langle d_{x^{2}-y^{2}} |\mathcal{H}| p_{x}^{\mathrm{in}} \rangle $ & $\myred{ 0.80 }$	                     &  $ \langle d_{xy} |\mathcal{H}| p_{x}^{\mathrm{in}} \rangle$ &  $\myred{ -0.79 }$  \\
$ \langle d_{xy} |\mathcal{H}| p_{x}^{\mathrm{in}} \rangle $          &  $\myred{ 0.80 }$  & $ \langle d_{xy} |\mathcal{H}| p_{x}^{\mathrm{in}} \rangle $ & $\myred{ 0.80 }$                                 &  $ \langle d_{x^{2}-y^{2}} |\mathcal{H}| p_{y}^{\mathrm{in}} \rangle$ &  $\myred{ -0.93 }$  \\
$ \langle d_{xz} |\mathcal{H}| p_{z}^{\mathrm{in}} \rangle $ 	      &  $0.62$  & $ \langle d_{xz} |\mathcal{H}| p_{z}^{\mathrm{in}} \rangle $ & $0.63$   		                             &  $ \langle d_{yz} |\mathcal{H}| p_{z}^{\mathrm{out}} \rangle$ &  $-0.73$  \\
$ \langle d_{x^{2}-y^{2}} |\mathcal{H}| p_{y}^{\mathrm{in}} \rangle $ &  $\myred{ -0.93 }$  & $ \langle d_{x^{2}-y^{2}} |\mathcal{H}| p_{y}^{\mathrm{in}} \rangle $ & $\myred{ -0.93 }$                      &  $ \langle d_{x^{2}-y^{2}} |\mathcal{H}| p_{z}^{\mathrm{out}} \rangle$ &  $0.55$  \\
$ \langle d_{yz} |\mathcal{H}| p_{z}^{\mathrm{out}} \rangle $         &  $0.73$  & $ \langle d_{yz} |\mathcal{H}| p_{z}^{\mathrm{out}} \rangle $ & $0.73$                   &  $ \langle d_{z^{2}} |\mathcal{H}| p_{y}^{\mathrm{out}} \rangle$ &  $0.59$  \\
$ \langle d_{x^{2}-y^{2}} |\mathcal{H}| p_{z}^{\mathrm{out}} \rangle $ &  $0.54$  & $ \langle d_{x^{2}-y^{2}} |\mathcal{H}| p_{z}^{\mathrm{out}} \rangle $ & $0.55$                        &  $ \langle p_{x}^{\mathrm{in}} |\mathcal{H}| p_{z}^{\mathrm{out}} \rangle$ &  $0.58$  \\
$ \langle d_{z^{2}} |\mathcal{H}| p_{y}^{\mathrm{out}} \rangle $       &  $0.58$  & $ \langle d_{z^{2}} |\mathcal{H}| p_{y}^{\mathrm{out}} \rangle $ & $0.58$                          	                     &  & \\  
$ \langle p_{x}^{\mathrm{out}} |\mathcal{H}| p_{x}^{\mathrm{out}} \rangle $ &  $\myred{ -0.84 }$  & $ \langle p_{x}^{\mathrm{out}} |\mathcal{H}| p_{x}^{\mathrm{out}} \rangle $ & $\myred{ -0.84 }$          &  & \\
$ \langle p_{y}^{\mathrm{out}} |\mathcal{H}| p_{y}^{\mathrm{out}} \rangle $ &  $\myred{ -1.41 }$  & $ \langle p_{y}^{\mathrm{out}} |\mathcal{H}| p_{y}^{\mathrm{out}} \rangle $ & $\myred{ -1.41 }$          &  & \\
$ \langle p_{x}^{\mathrm{in}} |\mathcal{H}| p_{z}^{\mathrm{out}} \rangle $ &  $0.58$  &  $ \langle p_{x}^{\mathrm{in}} |\mathcal{H}| p_{z}^{\mathrm{out}} \rangle $ & $0.58$                                 &  & \\
    
\hline
\hline
\end{tabular*}
\caption{Relevant hopping integrals (in eV) obtained from maximally-localized Wannier functions\cite{wannier90} for AV$_{3}$Sb$_{5}$ (A = K, Rb, Cs). For each material, the first column provides the matrix element between the Wannier functions ($\phi_{\alpha}$ and $\phi_{\beta}$) of the form $\langle \phi_{\alpha} | \mathcal{H} | \phi_{\beta} \rangle $, where $\alpha \neq \beta$ and $\mathcal{H}$ is the Wannier Hamiltonian. The second column gives the corresponding hopping integral. The hoppings with the largest magnitude are denoted in red. The table is organized in the following order: $d$-$d$, $d$-$p^{\mathrm{in}}$, $d$-$p^{\mathrm{out}}$, and $p$-$p$, where $p^{\mathrm{in}}$ ($p^{\mathrm{out}}$) refers to $p$ orbitals on the $\Sbin$ ($\Sbout$) sites.}
\label{tab:hoppings}
\end{table*}

\section{\label{sec:es_press_dope}Electronic structure upon applied pressure and doping}

\begin{figure}[ht]
\centering
\includegraphics[width=0.5\columnwidth]{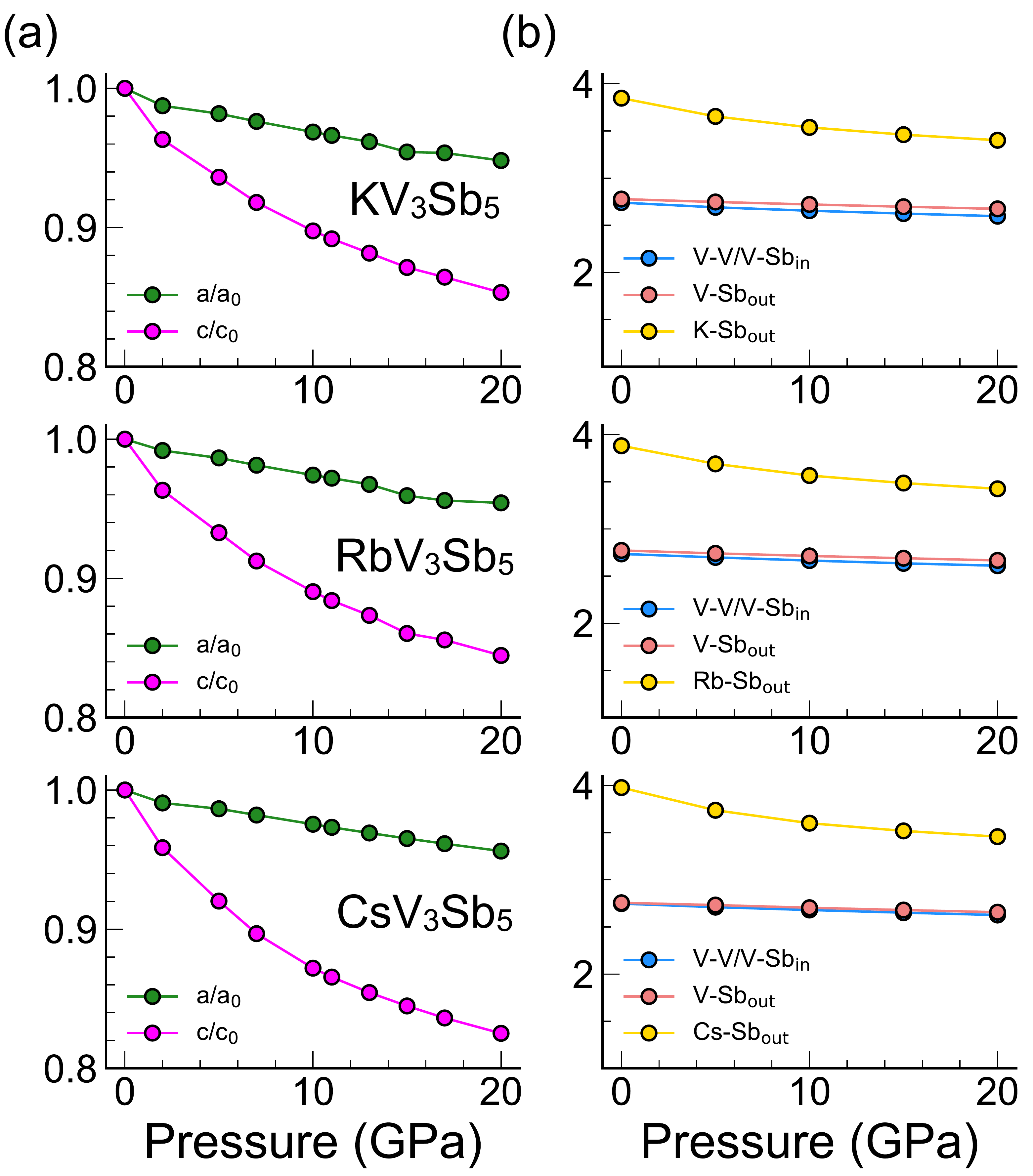}
\caption{Evolution upon applied pressure of (a) $a$ and $c$ lattice parameters (relative to their ambient pressure values $a_0$ and $c_0$), and (b) interatomic distances between V-V, V-$\Sbin$, V-$\Sbout$, and A-$\Sbout$ atoms (in units of \AA{})}
\label{fig:press_distance}
\end{figure}

\textit{Pressure.} In Fig. \ref{fig:press_distance}(a), we plot the change in lattice parameters and nearest-neighbor distances with applied pressure for all three AV$_3$Sb$_5$ compounds. As discussed in the main text, we find that the in-plane lattice parameter changes linearly with a maximum change of $\sim$ 3\% at the highest pressure. However, the out-of-plane lattice parameter exhibits a low ($\lesssim 8.5$ GPa) and high ($\gtrsim 8.5$ GPa) pressure regime where the lattice parameter changes at different rates with a much higher overall change of $\sim$ 15\% at the highest pressure calculated. From this data, we can calculate the ratio $c/a$, which has been shown to exhibit a kink near the termination of the first superconducting dome for the Cs compound \cite{zhang2021pressure}. In the main text, we show that this kink exists in the K and Rb compounds as well and qualitatively matches the termination of the superconducting dome in these materials. 

Concerning changes in the nearest-neighbor distances, we find that applying pressure (even up to 20 GPa) only has a minor effect on the nearest-neighbor distances between the V and Sb atoms (both in- and out-of-plane) (see Fig. \ref{fig:press_distance}(b)). In contrast, there is a much more dramatic change (nearly 0.5 \AA{} at 20 GPa) in the A-$\Sbout$ distance. This scenario supports that the V$_{3}$Sb$_{5}$ slabs are rigid structural layers that are only weakly coupled to the interstitial A cations. The evolution of the A-$\Sbout$ distance compared to all other nearest-neighbor bond lengths agrees with experimental data for the Cs compound \cite{tsirlin2021anisotropic}, and we show here that these trends can be extended to the K and Rb compounds. 
\begin{figure*}[ht]
	\centering
	\includegraphics[width=\columnwidth]{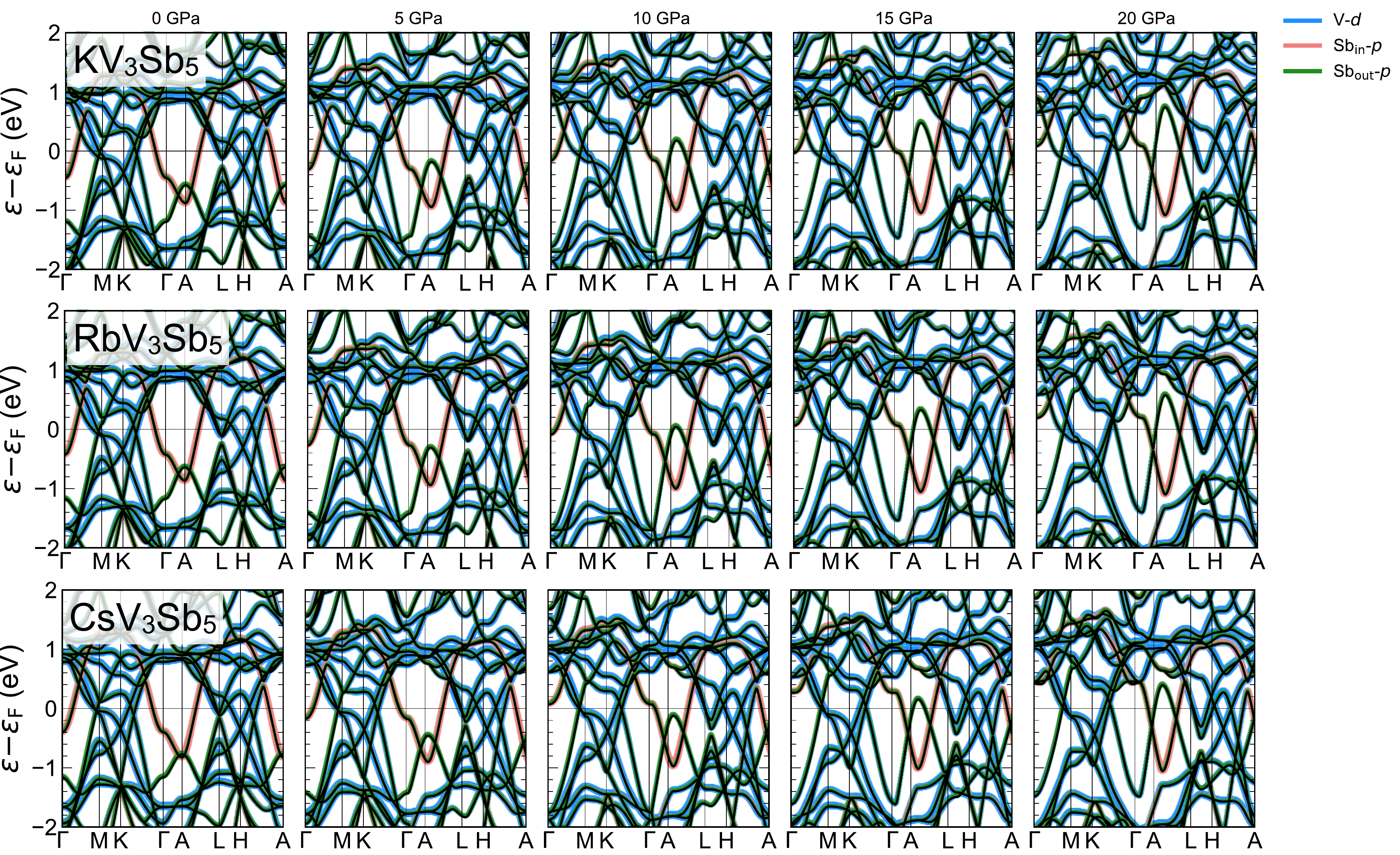}
	\caption{Evolution of the band structure under pressure for AV$_{3}$Sb$_{5}$: A = K (top), A = Rb (middle), and A = Cs (bottom). The band character for V-$d$, $\Sbin$-$p$, and $\Sbout$-$p$ is also shown.}
	\label{fig:pressure}
\end{figure*}

 In Fig. \ref{fig:pressure}, we provide the evolution of the band dispersions around the Fermi level with applied external pressure. The V-$d$ bands remain essentially the same as the applied pressure is increased. The vHss change relative to the Fermi level, but the dispersions and orbital content of the V-$d$ bands remain essentially unaffected. In contrast, the Sb-$p$ bands experience a major reconstruction.

\begin{figure}[ht]
    \centering
    \includegraphics[width=0.75\columnwidth]{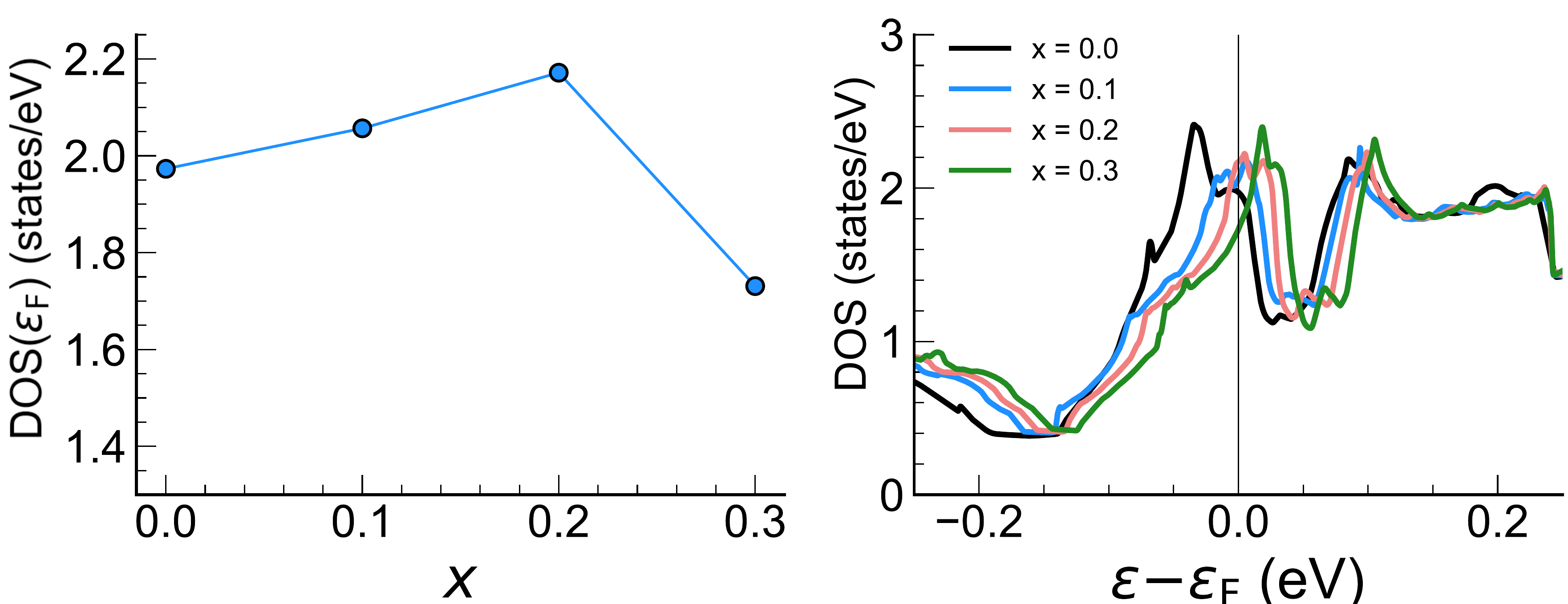}
    \caption{(a) Total density of states at $\eF$  and (b) total density of states within a small window around $\eF$ for CsV$_3$Sb$_5$ with respect to the doping level ($x$).}
    \label{fig:cs_dos}
\end{figure}

\textit{Doping.} Fig. \ref{fig:cs_dos} shows the total density of states (DOS) at and around the $\eF$ for the Cs compound upon hole doping. The increase in total DOS with doping (up to $x=0.2$) matches the increase in the superconducting critical temperature observed with hole doping in Ref. \onlinecite{song2021competition}. The decrease in total DOS at $x=0.3$ corresponds to the large DOS associated with vHs1 shifting above the Fermi level.

\end{document}